\documentclass{article}
\usepackage[left=1in, right=1in, top=1in, bottom=1in]{geometry}
\usepackage[numbers]{natbib}
\usepackage{amsmath}
\usepackage{mathtools}
\usepackage{subcaption}
\usepackage{caption}
\usepackage{graphicx}
\usepackage{amsthm}
\usepackage{amsfonts}
\usepackage{bbm}
\usepackage{blindtext}
\usepackage{placeins}
\usepackage{titlesec}
\usepackage{listings}
\usepackage{xcolor}
\usepackage[colorlinks]{hyperref}
\usepackage[colorinlistoftodos]{todonotes}
\usepackage{cleveref}
\crefname{equation}{Eq}{Eqs} 
\crefrangelabelformat{equation}{(#3#1#4--#5#2#6)}

\theoremstyle{definition}

\newcommand{\appendixsectionformat}{%
    \titleformat{\section}{\normalfont\Large\bfseries}{Appendix \thesection}{1em}{}}

\usepackage{parskip} 
\usepackage[most]{tcolorbox}

\usepackage{graphicx}
\setkeys{Gin}{keepaspectratio}

\usepackage{caption}

\usepackage{float}
\usepackage{xcolor} 
\usepackage{enumerate} 
\usepackage{geometry} 
\usepackage{amsmath} 
\usepackage{amssymb} 
\usepackage{textcomp} 
\AtBeginDocument{%
    \def\PYZsq{\textquotesingle}
}
\usepackage{upquote} 
\usepackage{eurosym} 

\usepackage{iftex}
\ifPDFTeX
    \usepackage[T1]{fontenc}
    \IfFileExists{alphabeta.sty}{
          \usepackage{alphabeta}
      }{
          \usepackage[mathletters]{ucs}
          \usepackage[utf8x]{inputenc}
      }
\else
    \usepackage{fontspec}
    \usepackage{unicode-math}
\fi

\usepackage{fancyvrb} 
\usepackage{grffile} 
\makeatletter 
\@ifpackagelater{grffile}{2019/11/01}
{
}
{
  \def\Gread@@xetex#1{%
    \IfFileExists{"\Gin@base".bb}%
    {\Gread@eps{\Gin@base.bb}}%
    {\Gread@@xetex@aux#1}%
  }
}
\makeatother
\usepackage[Export]{adjustbox} 
\adjustboxset{max size={0.9\linewidth}{0.9\paperheight}}

\usepackage{hyperref}
\usepackage{titling}
\usepackage{longtable} 
\usepackage{booktabs}  
\usepackage{array}     
\usepackage{calc}      
\usepackage[inline]{enumitem} 
\usepackage[normalem]{ulem} 
\usepackage{soul}      
\usepackage{mathrsfs}

\definecolor{urlcolor}{rgb}{0,.145,.698}
\definecolor{linkcolor}{rgb}{.71,0.21,0.01}
\definecolor{citecolor}{rgb}{.12,.54,.11}

\definecolor{ansi-black}{HTML}{3E424D}
\definecolor{ansi-black-intense}{HTML}{282C36}
\definecolor{ansi-red}{HTML}{E75C58}
\definecolor{ansi-red-intense}{HTML}{B22B31}
\definecolor{ansi-green}{HTML}{00A250}
\definecolor{ansi-green-intense}{HTML}{007427}
\definecolor{ansi-yellow}{HTML}{DDB62B}
\definecolor{ansi-yellow-intense}{HTML}{B27D12}
\definecolor{ansi-blue}{HTML}{208FFB}
\definecolor{ansi-blue-intense}{HTML}{0065CA}
\definecolor{ansi-magenta}{HTML}{D160C4}
\definecolor{ansi-magenta-intense}{HTML}{A03196}
\definecolor{ansi-cyan}{HTML}{60C6C8}
\definecolor{ansi-cyan-intense}{HTML}{258F8F}
\definecolor{ansi-white}{HTML}{C5C1B4}
\definecolor{ansi-white-intense}{HTML}{A1A6B2}
\definecolor{ansi-default-inverse-fg}{HTML}{FFFFFF}
\definecolor{ansi-default-inverse-bg}{HTML}{000000}

\definecolor{outerrorbackground}{HTML}{FFDFDF}

\DefineVerbatimEnvironment{Highlighting}{Verbatim}{commandchars=\\\{\}}

\let\Oldtex\TeX
\let\Oldlatex\LaTeX
\renewcommand{\TeX}{\textrm{\Oldtex}}
\renewcommand{\LaTeX}{\textrm{\Oldlatex}}
\title{garden-city}

\makeatletter
\def\PY@reset{\let\PY@it=\relax \let\PY@bf=\relax%
\let\PY@ul=\relax \let\PY@tc=\relax%
\let\PY@bc=\relax \let\PY@ff=\relax}
\def\PY@tok#1{\csname PY@tok@#1\endcsname}
\def\PY@toks#1+{\ifx\relax#1\empty\else%
\PY@tok{#1}\expandafter\PY@toks\fi}
\def\PY@do#1{\PY@bc{\PY@tc{\PY@ul{%
\PY@it{\PY@bf{\PY@ff{#1}}}}}}}
\def\PY#1#2{\PY@reset\PY@toks#1+\relax+\PY@do{#2}}

\@namedef{PY@tok@w}{\def\PY@tc##1{\textcolor[rgb]{0.73,0.73,0.73}{##1}}}
\@namedef{PY@tok@c}{\let\PY@it=\textit\def\PY@tc##1{\textcolor[rgb]{0.24,0.48,0.48}{##1}}}
\@namedef{PY@tok@cp}{\def\PY@tc##1{\textcolor[rgb]{0.61,0.40,0.00}{##1}}}
\@namedef{PY@tok@k}{\let\PY@bf=\textbf\def\PY@tc##1{\textcolor[rgb]{0.00,0.50,0.00}{##1}}}
\@namedef{PY@tok@kp}{\def\PY@tc##1{\textcolor[rgb]{0.00,0.50,0.00}{##1}}}
\@namedef{PY@tok@kt}{\def\PY@tc##1{\textcolor[rgb]{0.69,0.00,0.25}{##1}}}
\@namedef{PY@tok@o}{\def\PY@tc##1{\textcolor[rgb]{0.40,0.40,0.40}{##1}}}
\@namedef{PY@tok@ow}{\let\PY@bf=\textbf\def\PY@tc##1{\textcolor[rgb]{0.67,0.13,1.00}{##1}}}
\@namedef{PY@tok@nb}{\def\PY@tc##1{\textcolor[rgb]{0.00,0.50,0.00}{##1}}}
\@namedef{PY@tok@nf}{\def\PY@tc##1{\textcolor[rgb]{0.00,0.00,1.00}{##1}}}
\@namedef{PY@tok@nc}{\let\PY@bf=\textbf\def\PY@tc##1{\textcolor[rgb]{0.00,0.00,1.00}{##1}}}
\@namedef{PY@tok@nn}{\let\PY@bf=\textbf\def\PY@tc##1{\textcolor[rgb]{0.00,0.00,1.00}{##1}}}
\@namedef{PY@tok@ne}{\let\PY@bf=\textbf\def\PY@tc##1{\textcolor[rgb]{0.80,0.25,0.22}{##1}}}
\@namedef{PY@tok@nv}{\def\PY@tc##1{\textcolor[rgb]{0.10,0.09,0.49}{##1}}}
\@namedef{PY@tok@no}{\def\PY@tc##1{\textcolor[rgb]{0.53,0.00,0.00}{##1}}}
\@namedef{PY@tok@nl}{\def\PY@tc##1{\textcolor[rgb]{0.46,0.46,0.00}{##1}}}
\@namedef{PY@tok@ni}{\let\PY@bf=\textbf\def\PY@tc##1{\textcolor[rgb]{0.44,0.44,0.44}{##1}}}
\@namedef{PY@tok@na}{\def\PY@tc##1{\textcolor[rgb]{0.41,0.47,0.13}{##1}}}
\@namedef{PY@tok@nt}{\let\PY@bf=\textbf\def\PY@tc##1{\textcolor[rgb]{0.00,0.50,0.00}{##1}}}
\@namedef{PY@tok@nd}{\def\PY@tc##1{\textcolor[rgb]{0.67,0.13,1.00}{##1}}}
\@namedef{PY@tok@s}{\def\PY@tc##1{\textcolor[rgb]{0.73,0.13,0.13}{##1}}}
\@namedef{PY@tok@sd}{\let\PY@it=\textit\def\PY@tc##1{\textcolor[rgb]{0.73,0.13,0.13}{##1}}}
\@namedef{PY@tok@si}{\let\PY@bf=\textbf\def\PY@tc##1{\textcolor[rgb]{0.64,0.35,0.47}{##1}}}
\@namedef{PY@tok@se}{\let\PY@bf=\textbf\def\PY@tc##1{\textcolor[rgb]{0.67,0.36,0.12}{##1}}}
\@namedef{PY@tok@sr}{\def\PY@tc##1{\textcolor[rgb]{0.64,0.35,0.47}{##1}}}
\@namedef{PY@tok@ss}{\def\PY@tc##1{\textcolor[rgb]{0.10,0.09,0.49}{##1}}}
\@namedef{PY@tok@sx}{\def\PY@tc##1{\textcolor[rgb]{0.00,0.50,0.00}{##1}}}
\@namedef{PY@tok@m}{\def\PY@tc##1{\textcolor[rgb]{0.40,0.40,0.40}{##1}}}
\@namedef{PY@tok@gh}{\let\PY@bf=\textbf\def\PY@tc##1{\textcolor[rgb]{0.00,0.00,0.50}{##1}}}
\@namedef{PY@tok@gu}{\let\PY@bf=\textbf\def\PY@tc##1{\textcolor[rgb]{0.50,0.00,0.50}{##1}}}
\@namedef{PY@tok@gd}{\def\PY@tc##1{\textcolor[rgb]{0.63,0.00,0.00}{##1}}}
\@namedef{PY@tok@gi}{\def\PY@tc##1{\textcolor[rgb]{0.00,0.52,0.00}{##1}}}
\@namedef{PY@tok@gr}{\def\PY@tc##1{\textcolor[rgb]{0.89,0.00,0.00}{##1}}}
\@namedef{PY@tok@ge}{\let\PY@it=\textit}
\@namedef{PY@tok@gs}{\let\PY@bf=\textbf}
\@namedef{PY@tok@ges}{\let\PY@bf=\textbf\let\PY@it=\textit}
\@namedef{PY@tok@gp}{\let\PY@bf=\textbf\def\PY@tc##1{\textcolor[rgb]{0.00,0.00,0.50}{##1}}}
\@namedef{PY@tok@go}{\def\PY@tc##1{\textcolor[rgb]{0.44,0.44,0.44}{##1}}}
\@namedef{PY@tok@gt}{\def\PY@tc##1{\textcolor[rgb]{0.00,0.27,0.87}{##1}}}
\@namedef{PY@tok@err}{\def\PY@bc##1{{\setlength{\fboxsep}{\string -\fboxrule}\fcolorbox[rgb]{1.00,0.00,0.00}{1,1,1}{\strut ##1}}}}
\@namedef{PY@tok@kc}{\let\PY@bf=\textbf\def\PY@tc##1{\textcolor[rgb]{0.00,0.50,0.00}{##1}}}
\@namedef{PY@tok@kd}{\let\PY@bf=\textbf\def\PY@tc##1{\textcolor[rgb]{0.00,0.50,0.00}{##1}}}
\@namedef{PY@tok@kn}{\let\PY@bf=\textbf\def\PY@tc##1{\textcolor[rgb]{0.00,0.50,0.00}{##1}}}
\@namedef{PY@tok@kr}{\let\PY@bf=\textbf\def\PY@tc##1{\textcolor[rgb]{0.00,0.50,0.00}{##1}}}
\@namedef{PY@tok@bp}{\def\PY@tc##1{\textcolor[rgb]{0.00,0.50,0.00}{##1}}}
\@namedef{PY@tok@fm}{\def\PY@tc##1{\textcolor[rgb]{0.00,0.00,1.00}{##1}}}
\@namedef{PY@tok@vc}{\def\PY@tc##1{\textcolor[rgb]{0.10,0.09,0.49}{##1}}}
\@namedef{PY@tok@vg}{\def\PY@tc##1{\textcolor[rgb]{0.10,0.09,0.49}{##1}}}
\@namedef{PY@tok@vi}{\def\PY@tc##1{\textcolor[rgb]{0.10,0.09,0.49}{##1}}}
\@namedef{PY@tok@vm}{\def\PY@tc##1{\textcolor[rgb]{0.10,0.09,0.49}{##1}}}
\@namedef{PY@tok@sa}{\def\PY@tc##1{\textcolor[rgb]{0.73,0.13,0.13}{##1}}}
\@namedef{PY@tok@sb}{\def\PY@tc##1{\textcolor[rgb]{0.73,0.13,0.13}{##1}}}
\@namedef{PY@tok@sc}{\def\PY@tc##1{\textcolor[rgb]{0.73,0.13,0.13}{##1}}}
\@namedef{PY@tok@dl}{\def\PY@tc##1{\textcolor[rgb]{0.73,0.13,0.13}{##1}}}
\@namedef{PY@tok@s2}{\def\PY@tc##1{\textcolor[rgb]{0.73,0.13,0.13}{##1}}}
\@namedef{PY@tok@sh}{\def\PY@tc##1{\textcolor[rgb]{0.73,0.13,0.13}{##1}}}
\@namedef{PY@tok@s1}{\def\PY@tc##1{\textcolor[rgb]{0.73,0.13,0.13}{##1}}}
\@namedef{PY@tok@mb}{\def\PY@tc##1{\textcolor[rgb]{0.40,0.40,0.40}{##1}}}
\@namedef{PY@tok@mf}{\def\PY@tc##1{\textcolor[rgb]{0.40,0.40,0.40}{##1}}}
\@namedef{PY@tok@mh}{\def\PY@tc##1{\textcolor[rgb]{0.40,0.40,0.40}{##1}}}
\@namedef{PY@tok@mi}{\def\PY@tc##1{\textcolor[rgb]{0.40,0.40,0.40}{##1}}}
\@namedef{PY@tok@il}{\def\PY@tc##1{\textcolor[rgb]{0.40,0.40,0.40}{##1}}}
\@namedef{PY@tok@mo}{\def\PY@tc##1{\textcolor[rgb]{0.40,0.40,0.40}{##1}}}
\@namedef{PY@tok@ch}{\let\PY@it=\textit\def\PY@tc##1{\textcolor[rgb]{0.24,0.48,0.48}{##1}}}
\@namedef{PY@tok@cm}{\let\PY@it=\textit\def\PY@tc##1{\textcolor[rgb]{0.24,0.48,0.48}{##1}}}
\@namedef{PY@tok@cpf}{\let\PY@it=\textit\def\PY@tc##1{\textcolor[rgb]{0.24,0.48,0.48}{##1}}}
\@namedef{PY@tok@c1}{\let\PY@it=\textit\def\PY@tc##1{\textcolor[rgb]{0.24,0.48,0.48}{##1}}}
\@namedef{PY@tok@cs}{\let\PY@it=\textit\def\PY@tc##1{\textcolor[rgb]{0.24,0.48,0.48}{##1}}}

\def\PYZus{\char`\_}
\def\PYZob{\char`\{}
\def\PYZcb{\char`\}}

\def\PYZsh{\char`\#}

\def\PYZhy{\char`\-}
\def\PYZsq{\char`\'}
\def\PYZdq{\char`\"}

\makeatother

\makeatletter
    \newbox\Wrappedcontinuationbox
    \newbox\Wrappedvisiblespacebox
    \newcommand*\Wrappedvisiblespace {\textcolor{red}{\textvisiblespace}}
    \newcommand*\Wrappedcontinuationsymbol {\textcolor{red}{\llap{\tiny$\m@th\hookrightarrow$}}}
    \newcommand*\Wrappedcontinuationindent {3ex }
    \newcommand*\Wrappedafterbreak {\kern\Wrappedcontinuationindent\copy\Wrappedcontinuationbox}
    \newcommand*\Wrappedbreaksatspecials {%
        \def\PYGZus{\discretionary{\char`\_}{\Wrappedafterbreak}{\char`\_}}%
        \def\PYGZob{\discretionary{}{\Wrappedafterbreak\char`\{}{\char`\{}}%
        \def\PYGZcb{\discretionary{\char`\}}{\Wrappedafterbreak}{\char`\}}}%
        \def\PYGZca{\discretionary{\char`\^}{\Wrappedafterbreak}{\char`\^}}%
        \def\PYGZam{\discretionary{\char`\&}{\Wrappedafterbreak}{\char`\&}}%
        \def\PYGZlt{\discretionary{}{\Wrappedafterbreak\char`\<}{\char`\<}}%
        \def\PYGZgt{\discretionary{\char`\>}{\Wrappedafterbreak}{\char`\>}}%
        \def\PYGZsh{\discretionary{}{\Wrappedafterbreak\char`\#}{\char`\#}}%
        \def\PYGZpc{\discretionary{}{\Wrappedafterbreak\char`\%}{\char`\%}}%
        \def\PYGZdl{\discretionary{}{\Wrappedafterbreak\char`\$}{\char`\$}}%
        \def\PYGZhy{\discretionary{\char`\-}{\Wrappedafterbreak}{\char`\-}}%
        \def\PYGZsq{\discretionary{}{\Wrappedafterbreak\textquotesingle}{\textquotesingle}}%
        \def\PYGZdq{\discretionary{}{\Wrappedafterbreak\char`\"}{\char`\"}}%
        \def\PYGZti{\discretionary{\char`\~}{\Wrappedafterbreak}{\char`\~}}%
    }
    \newcommand*\Wrappedbreaksatpunct {%
        \lccode`\~`\.\lowercase{\def~}{\discretionary{\hbox{\char`\.}}{\Wrappedafterbreak}{\hbox{\char`\.}}}%
        \lccode`\~`\,\lowercase{\def~}{\discretionary{\hbox{\char`\,}}{\Wrappedafterbreak}{\hbox{\char`\,}}}%
        \lccode`\~`\;\lowercase{\def~}{\discretionary{\hbox{\char`\;}}{\Wrappedafterbreak}{\hbox{\char`\;}}}%
        \lccode`\~`\:\lowercase{\def~}{\discretionary{\hbox{\char`\:}}{\Wrappedafterbreak}{\hbox{\char`\:}}}%
        \lccode`\~`\?\lowercase{\def~}{\discretionary{\hbox{\char`\?}}{\Wrappedafterbreak}{\hbox{\char`\?}}}%
        \lccode`\~`\!\lowercase{\def~}{\discretionary{\hbox{\char`\!}}{\Wrappedafterbreak}{\hbox{\char`\!}}}%
        \lccode`\~`\/\lowercase{\def~}{\discretionary{\hbox{\char`\/}}{\Wrappedafterbreak}{\hbox{\char`\/}}}%
        \catcode`\.\active
        \catcode`\,\active
        \catcode`\;\active
        \catcode`\:\active
        \catcode`\?\active
        \catcode`\!\active
        \catcode`\/\active
        \lccode`\~`\~
    }
\makeatother

\let\OriginalVerbatim=\Verbatim
\makeatletter
\renewcommand{\Verbatim}[1][1]{%
    \sbox\Wrappedcontinuationbox {\Wrappedcontinuationsymbol}%
    \sbox\Wrappedvisiblespacebox {\FV@SetupFont\Wrappedvisiblespace}%
    \def\FancyVerbFormatLine ##1{\hsize\linewidth
        \vtop{\raggedright\hyphenpenalty\z@\exhyphenpenalty\z@
            \doublehyphendemerits\z@\finalhyphendemerits\z@
            \strut ##1\strut}%
    }%
    \def\FV@Space {%
        \nobreak\hskip\z@ plus\fontdimen3\font minus\fontdimen4\font
        \discretionary{\copy\Wrappedvisiblespacebox}{\Wrappedafterbreak}
        {\kern\fontdimen2\font}%
    }%

    \Wrappedbreaksatspecials
    \OriginalVerbatim[#1,codes*=\Wrappedbreaksatpunct]%
}
\makeatother

\definecolor{incolor}{HTML}{303F9F}
\definecolor{outcolor}{HTML}{D84315}
\definecolor{cellborder}{HTML}{CFCFCF}
\definecolor{cellbackground}{HTML}{F7F7F7}

\makeatletter
\newcommand{\boxspacing}{\kern\kvtcb@left@rule\kern\kvtcb@boxsep}
\makeatother

\hypersetup{
  breaklinks=true,  
  colorlinks=true,
  urlcolor=urlcolor,
  linkcolor=linkcolor,
  citecolor=citecolor,
  }

\title{Garden city: A synthetic dataset and sandbox environment for analysis of pre-processing algorithms for GPS human mobility data}

\author{Thomas H. Li, Francisco Barreras}
\date{}

\begin{document}

\maketitle
\begin{abstract}
\noindent Human mobility datasets have seen increasing adoption in the past decade, enabling diverse applications that leverage the high precision of measured trajectories relative to other human mobility datasets. However, there are concerns about whether the high sparsity in some commercial datasets can introduce errors due to lack of robustness in processing algorithms, which could compromise the validity of downstream results. The scarcity of ``ground-truth'' data makes it particularly challenging to evaluate and calibrate these algorithms. To overcome these limitations and allow for an intermediate form of validation of common processing algorithms, we propose a synthetic trajectory simulator and sandbox environment meant to replicate the features of commercial datasets that could cause errors in such algorithms, and which can be used to compare algorithm outputs with ``ground-truth'' synthetic trajectories and mobility diaries. Our code is open-source and is publicly available alongside tutorial notebooks and sample datasets generated with it. 
    \end{abstract}
\section{Introduction}
\begin{figure}[!ht]
    \centering
    \includegraphics[width=\linewidth]{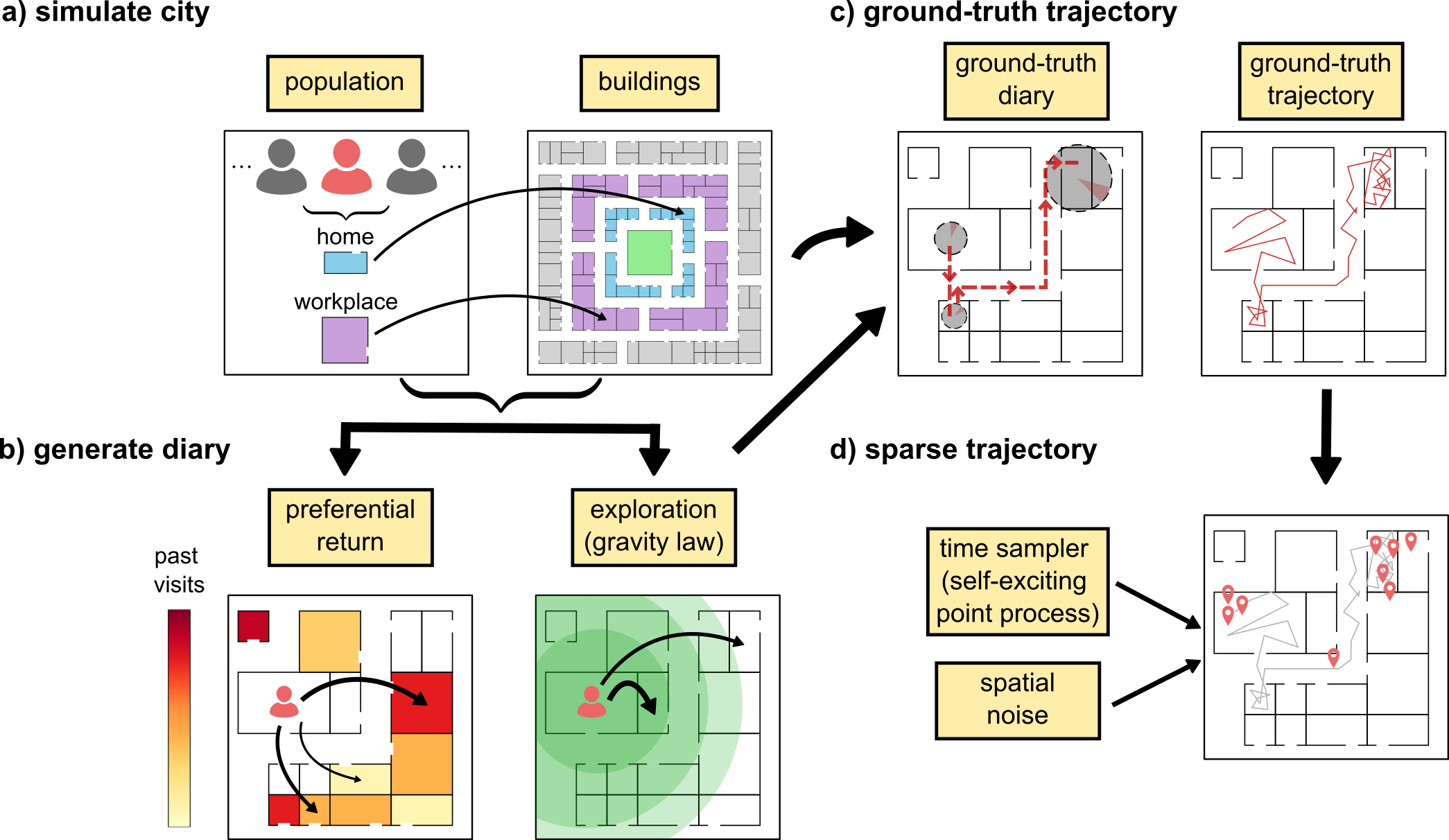}
    \caption{Outline of our sparse trajectory generator which proceeds in four stages. \textbf{a)} A city is generated from aligned building blocks, and a population of agents. \textbf{b)} A ``mobility diary'' can be generated using an exploration and preferential return (EPR) model. \textbf{c)} A complete ``ground-truth'' trajectory is generated combining a generated diary (``macro'' mobility) and a Brownian motion constrained to the current building's boundaries (``micro'' mobility). \textbf{d)} Finally, a sparse trajectory can be generated by sampling random times from the complete trajectory according to a self-exciting point process. Gaussian noise is added to the coordinates to simulate measurement error. }
    \label{fig:outline}
\end{figure}

Human mobility datasets, particularly GPS data collected from smartphone apps, have become indispensable in a wide range of applications such as epidemic modeling \cite{chang2021mobility, pepe2020covid, alessandretti2022human, couture2021jue}, disaster management \cite{song2014prediction}, studies on segregation and inequality \cite{moro2021mobility, gauvin2020gender}, transportation planning \cite{becker2017literature, cresswell2006move}, healthcare analytics \cite{pappalardo2016analytical}, and analyses of human behavior \cite{painter2021political, adjodah2021association}. Despite their widespread use, it has been noted that these commercial datasets often suffer from biases, high sparsity, and noise, which can introduce errors on trajectory pre-processing algorithms used in most of these applications \cite{barreras2024exciting}.

The processing pipeline from raw GPS data into mobility metrics or models typically involves applying algorithms for stop detection \cite{hariharan2004project, ross2021household, aslak2020infostop, yang2020tad}, home and work location attribution \cite{vanhoof2018assessing, couture2021jue}, trip detection and classification \cite{martin2023trackintel}, and contact estimation \cite{pepe2020covid}, among others. Many of these algorithms, which might perform well on complete regular data, might lack robustness to data issues like high sparsity or increased GPS noise from a dense urban environment. Furthermore, it has been suggested that the optimal parametrization of these algorithms could depend on heterogeneous features of user movement, signal quality, and environment \cite{barreras2024exciting}. Without access to ``ground-truth'' data, it becomes difficult to rigorously evaluate the accuracy and robustness of these algorithms under realistic conditions.

To address this gap, we introduce  ``Garden City'', an agent-based model designed to generate synthetic human mobility trajectories that serve as a proxy for ``ground-truth'' for the purpose of testing processing algorithms. Our model replicates key features of commercial GPS datasets, like noise, sparsity, and temporal clustering of pings into \emph{bursts}, enabling researchers to evaluate the parameter regimes in which algorithms are accurate and robust. 

The Garden City mobility model consists of three parts: a generated city which consists of rectangular buildings and rectilinear streets connecting them; a synthetic population endowed with an exploration and preferential return (EPR) mobility model that describes which locations are visited throughout the day; and a sparse trajectory sampler which takes into account the noise added by ``micro-level'' randomness in movement, the noise added by GPS sensors in smartphones, and, crucially, the temporal irregularity in ping collection characterized by bursts and long gaps of inactivity. In what remains of this section we give a brief overview of each of the parts of the simulation methods. 

\subsection{Synthetic city}

\begin{figure}[!ht]
    \centering
    \includegraphics[width=0.4\linewidth]{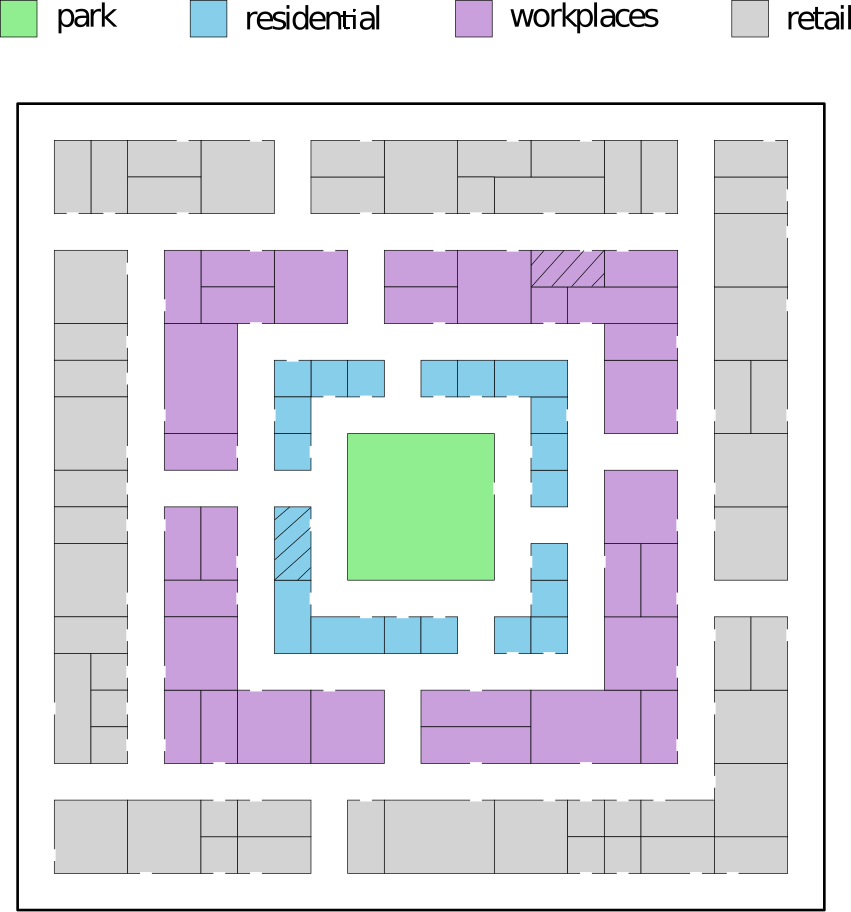}
    \caption{Example layout of a simulated city with a radial layout inspired by E. Howard's ideal ``garden city'' concept \cite{howard1898garden}. The city contains a park (the garden) at the center, with concentric rings of residential, workplace and retail locations. The shaded areas represent the home location and workplace location assigned to a given agent.}
    \label{fig:empty-city}
\end{figure}

The city in which our agents' trajectories unfold is represented as a collection of buildings which are rectangles on a grid made of building blocks. Each building is assigned a building type---e.g. homes, workplaces, and retail---as well as areas representing parks. Additionally, buildings have a ``door'' attribute, connecting them to the street---which we use during the trajectory simulation as the start and end points of trajectories inside the building. Figure \ref{fig:empty-city} demonstrates a possible layout, in which buildings are arranged in concentric rings, with a park in the center, following a ring of residences (blue), a ring of retail locations (purple), and, finally, an outer ring of workplaces (grey).\footnote{This layout is chosen as a nod to the design proposed by E. Howard of an ideal city in 1898 \cite{howard1898garden}.}

\subsection{Agents and mobility diaries} 

Agents are represented by a class that contains a ``work'' and ``home'' location, as well as a ``mobility diary'' which is represented with a discrete stochastic process in which agents move from building to building with transition probabilities depending on the time of the day and the previous locations visited. The specific values of these transitions are implemented using a variation of the Exploration and Preferential Return (EPR) model \cite{pappalardo2018data}, in which agents return to previous locations with a certain probability---proportional to the relative frequency of previous visits---or, otherwise, explore new a building with a probability decreasing with the square of the distance. The accuracy of the EPR model and its variants in capturing the distribution of locations visited and travel patterns has been demonstrated empirically \cite{song2010modelling, pappalardo2018data, alessandretti2018evidence}. Our mobility model is a modification of the EPR model described in \cite{pappalardo2016human}, in which the transition probabilities are biased towards certain building types depending on the time of the day, so as to capture the realistic circadian rhythms of individuals (e.g., forbidding individuals to ``return to the workplace'' in the middle of the night).

\subsection{Sparse trajectory generation}

Trajectories of arbitrary duration are sampled based on their circadian patterns using a two-step process. First we produce a ``ground-truth'' trajectory, which combines the ``macro'' trajectory dynamics given by the EPR model to choose the next building to be visited, with random ``micro'' trajectory dynamics based on Brownian motions that are constrained to remain inside the building being visited. Such ``micro'' dynamics can be tuned so that the walking speed varies across different building types---e.g., to model how agents are more likely to stay still when they are at their homes late at night, or at their workplaces.

The second step is to sample from the ``ground-truth'' trajectory generated in step one at random irregular times that model signal sparsity and clustering of pings into ``bursts'', as well as adding Gaussian noise to the location sampled to model measurement error from the GPS sensor. These random times are chosen using a self-exciting point process that can capture clustering of the pings into bursts.

\begin{figure}
\footnotesize
\begin{minipage}{0.08\textwidth}
\hspace{1pt}
\end{minipage}~
\begin{minipage}{0.8\textwidth}
    \input{snippet-building}
\end{minipage}
    \caption{Example code displaying the attributes of an instance of the class \texttt{Building} in a generated city.}
    \label{fig:building-attr}
\end{figure}

In what follows we describe these different components in details and provide examples. 

\section{Synthetic city generation}

We aim to create a flexible sandbox environment for urban simulation that enables researchers to explore and test spatial processing pipelines in a controlled, yet realistic, setting. The Garden City model is designed to replicate real-world urban environments, with the simplicity and convenience of being made up of building blocks. Because of the emphasis on studying the robustness of trajectory processing algorithms, the buildings are simplified to only include the features most likely to affect the algorithms' output---namely, variations in size, typical dwell times, and attributes modulating the speed at which agents move \emph{within} the buildings. A small toy example is shown in Figure \ref{fig:empty-city}, which might be sufficient to test many features of clustering algorithms. However, the code is flexible enough to model cities of arbitrary size and complexity, potentially approximating real cities as in \cite{zufle2023urban}. 

In Garden City, a \texttt{City} class is initialized as an $N \times M$ grid of blocks, where each block corresponds to a 15-by-15 meter square (roughly the size of an average bodega in New York City). Each block can either belong to an instance of the class \texttt{Building}, or it is assumed to be a \texttt{Street} otherwise. Users can add objects of the class \texttt{Building} to a \texttt{City} instance, specifying the building type and the composing blocks (alternatively, with a bounding box), as well as the \texttt{Street} block in which the building door will face. A unique building identifier can be specified, otherwise defaulting to an id combining the coordinates and building type.

To aid in generating \texttt{Agent} movements between buildings following the shortest paths, our code allows users to front-load the computation of the shortest path between every pair of \texttt{Street} blocks by running the \texttt{get\_street\_graph} method, which implements Dijkstra's algorithm in the adjacency graph of streets. 

\section{Agents' mobility diary generator as EPR model} \label{sec:diary-gen}

We simulate mobility diaries for our agents in our model using a stochastic process that aims to combine two concepts found in the literature. First, the concept of a typical circadian routine, wherein individuals' have a circadian life cycle based on their attributes and needs---such as being at their home, being at their workplace, conducting business in other workplaces, going to restaurants and retail, or visiting other agents' homes---a concept that has been explored in the mobility generators found in \cite{zufle2023urban}. Second, we incorporate the concept of individuals \emph{deviating} from their typical routine and exploring a variety of locations, which we implement by assigning specific transition probabilities between buildings (to be visited) resulting from an EPR model (e.g., see \cite{pappalardo2018data})---wherein individuals return to visited places with probabilities proportional to previous visit frequencies or, alternatively, they explore the city prioritizing nearby locations with a gravity law \cite{pappalardo2016human}.

To implement the notion of a circadian routine, we modify the EPR model by constraining the set of buildings that an agent is allowed to visit at each time (e.g., between hours 20:00 and 03:00 agents may only visit residential locations). This guarantees that, on average, individuals start and end their days at home, go to a workplace during office hours, and visit retail locations and parks before and after work, as well as during a midday lunch break. The parameters governing these circadian routines are tuneable; however, given the emphasis on evaluating trajectory processing algorithms, our focus is limited to generating a distribution of building dwell times that is consistent with a typical routine---long dwells at home and at work, with short visits to retail, park, and other businesses.\footnote{Another difference with the model in \cite{pappalardo2016human} is that visitation probabilities are not influenced by ``collective preferences'', since they are not immediately relevant to the evaluation of trajectory processing algorithms.} 

\begin{figure}
    \centering
    \includegraphics[width=0.8\linewidth]{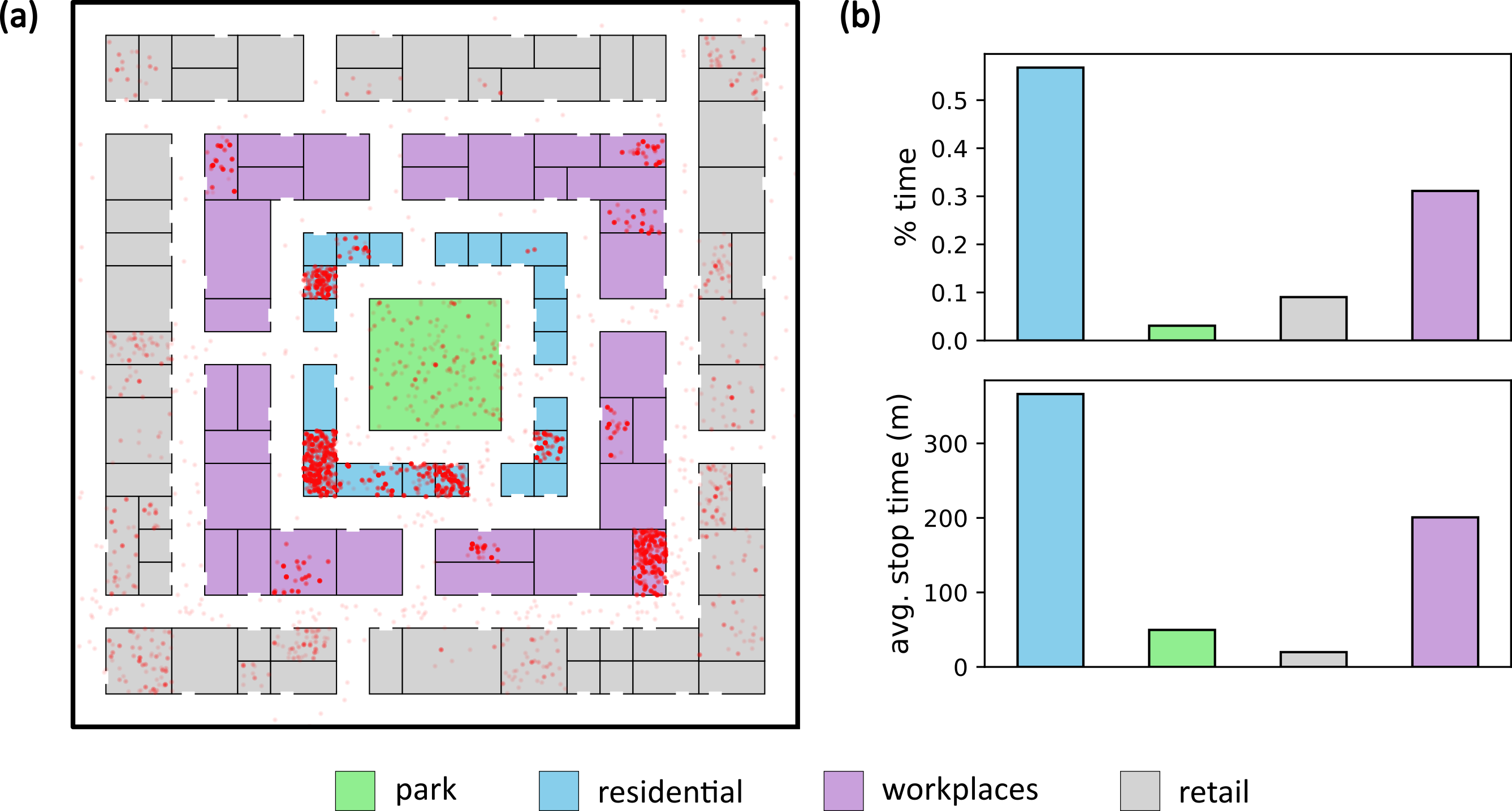}
    \caption{Visualization of a generated ``ground-truth'' trajectory and summary statistics of the corresponding travel diary of a simulated agent. Panel \textbf{(a)} approximates the frequency of locations visited by showing a scatter plot of points in the trajectory every minute for a week-long trajectory. Panel \textbf{(b)} provides a summary of the agent's diary, showing the percentage of time spent at each type of location as well as the average stop duration at each of the building types. }
    \label{fig:bob-in-the-city}
\end{figure}

Formally, we represent an agent's ``mobility diary'' generator as a discrete stochastic process with state variable $X_t$, with $t$ indexing time intervals of a fixed duration $\Delta$ (e.g. 15 minute intervals) up to time $T$. The state $X_t$ can take values in the set $\mathcal{L} \coloneqq \{ 1, 2, \dots, L\}$ indexing buildings. Furthermore, we define an auxiliary state vector $\mathbf{S}^{(t)}$ to represent the number of previous visits to buildings---that is, $S^{(t)}_{\ell}$ is the number of visits to building $\ell$ up to time $t$, which can be initialized with a non-zero vector $\mathbf{S}^{(0)}$. Additionally, let the set of unexplored buildings at time $t$ be $U^{(t)} \coloneqq \{\ell \in \mathcal{L} : S^{(t)}_{\ell} = 0 \}$, and let $V^{(t)} \coloneqq \mathcal{L} \setminus \left(U^{(t)} \cup \{X_t\} \right)$ be the set of visited buildings at time $t$ different than the current building. Furthermore, let $\mathbf{b}$ be a vector containing building types, so that $b_{\ell}$ is the type of building $\ell$. The stochastic process $\mathbf{X} \coloneqq \{X_1, X_2, \dots, X_T\}$ is parameterized by transition probabilities that we define in two steps. First, we generate a matrix of \emph{unconstrained} transition probabilities, $\mathbf{p}^{(t)}$, in which the transition from building $k$ to building $\ell$ at time $t$, namely $p_{k, \ell}^{(t)}$, is a function of $\mathbf{S}^{(t)}$, the visitation frequencies up to time $t$, and of the current location 
\begin{align}
    p_{k, \ell}^{(t)} \coloneqq P\left( X_{t+1} = \ell \mid X_t = k, \mathbf{S}^{(t)} = \mathbf{S} \right).
\end{align}
These unconstrained probabilities are chosen so as to satisfy the following requirements:
\begin{itemize}
    \item \textbf{There is a high probability of staying in the same location} from one time step to the next, so that stay durations at a specific \emph{building type} are approximately exponential with a prescribed mean. 
    \item \textbf{The probability of exploring new buildings decays as a power law} of the unique locations visited. That is, conditional on leaving the current building, the probabilities of visiting buildings in $U^{(t)}$ must, in total, be proportional to $\left(|V^{(t)}|\right)^{-\gamma}$.
    \item \textbf{Exploration follows a gravity law}. Formally, conditional on the next building being in $U^{(t)}$, then the probability of visiting building $\ell$ from building $k$ is inversely proportional to the squared distance between them, namely $r_{k,\ell}^2$. 
    \item \textbf{Preferential return probabilities are proportional to past visitation frequency}. That is, conditional on returning to a previously visited building, then probabilities of visiting are proportional to $\mathbf{S}^{(t)}$.
\end{itemize}
These requirements are summarized with the following equations
\begin{align} 
    p_{k, k}^{(t)} &= \kappa_{b_k} \label{eq:kk} \\ 
    \sum_{\ell \in U^{(t)}} p_{k, \ell}^{(t)} &= \rho (|V^{(t}|)^{-\gamma} \label{eq:exp} \\ 
    p_{k, \ell}^{(t)} &\sim \dfrac{1}{r_{k,\ell}^2}, \hspace{12pt} \text{for } \ell \in  U^{(t)}\label{eq:grav} \\ 
    p_{k, \ell}^{(t)} &\sim S^{(t)}_{\ell}, \hspace{12pt} \text{for } \ell \in  V^{(t)} \label{eq:freq}
\end{align}
Notice how the probability of ``jumping'' to a new location after $h$ time steps is given by $\left(1-\kappa_{b_k}\right)^h \sim e^{-\kappa_{b_k}h}$; thus, the stop time is approximately exponential with mean $1/\kappa_{b_k}$. Furthermore, the distance is computed as a \emph{Manhattan} distance using the adjacency graph of street blocks. 

\begin{figure}
\footnotesize
\begin{minipage}{0.08\textwidth}
\hspace{1pt}
\end{minipage}~
\begin{minipage}{0.8\textwidth}
    \input{snippet-diary-custom}
\end{minipage}
    \caption{Example code that generates an agent's ``mobility diary'' and ``ground-truth'' trajectory, from a custom destination diary passed by the user.}
    \label{fig:diary-snippet-custom}
\end{figure}

Finally, the final constrained transition probabilities, which model a circadian routine, are simply given by the conditional probabilities 
\[
    \overline{p}_{k, \ell}^{(t)} = \begin{cases} 
p_{k, \ell}^{(t)}/p_{k, A}^{(t)} , & \text{if } b_{\ell} \in A^{(t)}\\ 
0, & \text{otherwise} 
\end{cases}
\]
\noindent where $A^{(t)}$ is a set of allowed buildings at time $t$, and $p_{k, A}^{(t)}$ is the (unconstrained) probability of jumping to a building in $A^{(t)}$. Notice that none of the equations explicitly enforce a preference for staying home or at their workplace, rather, we achieve this behavior by initializing $\mathbf{S}^{(0)}$ with high values for the home location and work location. A small number of random locations can also be initialized with positive values to obtain a behavior closer to the asymptotic behavior, removing the need for a burnout period if the analysis is focused on ``regular'' mobility.  

\begin{figure}[hbt!]
\footnotesize
\begin{minipage}{0.08\textwidth}
\hspace{1pt}
\end{minipage}~
\begin{minipage}{0.8\textwidth}
    \input{snippet-diary-gen}
\end{minipage}
    \caption{Example code that generates an agent's ``mobility diary'' using the EPR model.}
    \label{fig:diary-snippet-gen}
\end{figure}

In practice, agents can be initialized with an initial location (defaulting to their home) and a ``mobility diary'' provided by the user in the form of a Pandas \texttt{DataFrame} or, alternatively, a diary can be generated by sampling from the EPR stochastic process given by \crefrange{eq:kk}{eq:freq}. This diary is stored in the agent's attribute \texttt{destination\_diary}, and represents a ``plan'' for the agent's trajectory. However, the sampled trajectory produces a different \emph{realized} diary, stored in the \texttt{diary} attribute, which takes into account the time traveling between locations, and which should be used as ``ground-truth''. Examples of these ``ground-truth'' diaries are shown in Figures \ref{fig:diary-snippet-custom} and  \ref{fig:diary-snippet-gen}, while details on the sampling of trajectories are specified in Section \ref{sec:traj-gen}. Panel (a) of Figure \ref{fig:bob-in-the-city} illustrates a generated complete trajectory for an example user for a period of one week, while panel (b) summarizes the total time and average stop time at each location type. 

\section{Sparse trajectory generation} \label{sec:traj-gen}

In this section, we give an overview of the methodology for simulating granular GPS trajectory data. Starting with an agent’s \texttt{destination\_diary}---which outlines their ``macro-level'' movement plan---we model the ``micro-level'' randomness in mobility, both within and between buildings, using constrained Brownian motions. A ``complete'' or ``ground-truth'' trajectory (e.g. minute-level) for an agent can be sampled from such Brownian motions, also generating a ``ground-truth mobility diary'' summarizing the ``macro-level'' user trajectory---specifically a table with stops and trips as shown in Figure \ref{fig:diary-snippet-gen}. The sparse GPS-like trajectory can then be obtained by sub-sampling the ``ground-truth'' trajectory at random times given by a self-exciting point process---which can be interpreted as ``sparsifying'' the complete trajectory, and then adding measurement error. In combination, this methodology produces a ``ground-truth mobility diary'', a ``complete'' trajectory, and a ``sparsified'' version of that trajectory, which can be used to test the accuracy and robustness of GPS human mobility data processing algorithms.

\subsection{``Micro'' dynamics: ``ground-truth'' trajectories from ``mobility diaries''}

Agents following a planned trajectory given by the \texttt{destination\_diary}---either provided by the user, or generated according to the model described in Section \ref{sec:diary-gen}---can either be \emph{inside} a building or traveling \emph{between} buildings. We model ``micro-level'' variance in agents' positions motivated by the impact that such variations can have on stop-detection algorithms---e.g. by increasing the roaming area within a location, or by producing positions close to the building boundaries which can result in (noisy) pings falling outside the building. We propose a simple model such that, when inside a building, an agent's position is constrained to the building’s boundaries, and such that there is a non-zero probability of the individual staying still at any instant. 

\begin{figure}
    \centering
    \includegraphics[width=0.6\linewidth]{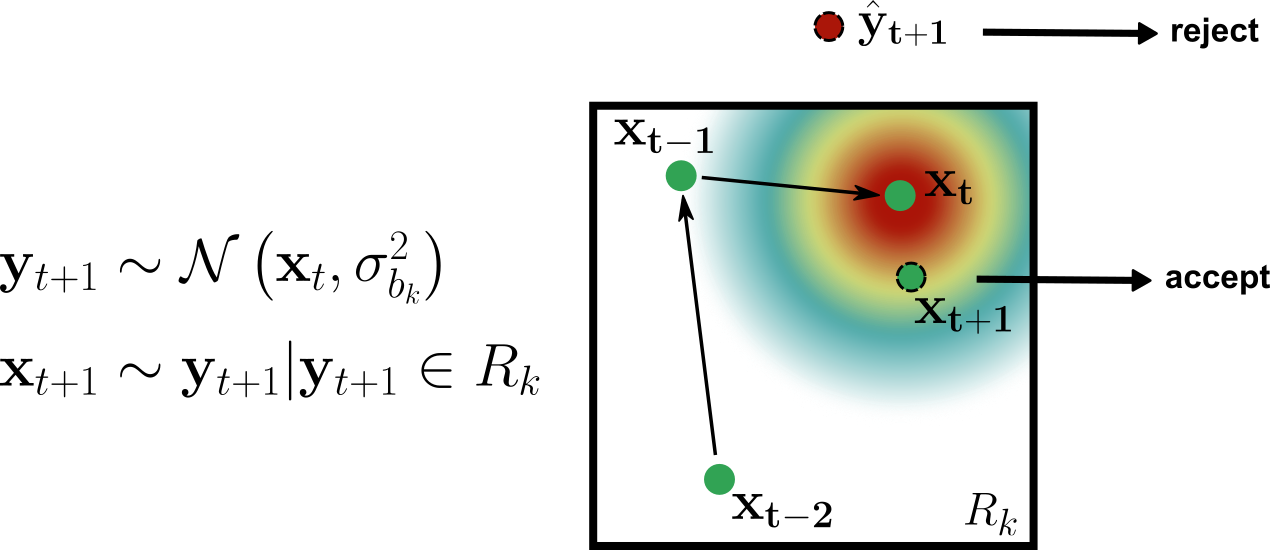}
    \caption{Sequence of samples from $\mathbf{x}_t$, which are constrained to be inside building $k$. Successive draws are of $\mathbf{y}_{t}$ can be sampled from a standard normal (with a point mass at $\mathbf{x}_t$), until a point falls inside the $R_k$, the region spanned by the building $k$.}
    \label{fig:constrained-brownian}
\end{figure}

Formally, the 2-dimensional variable $\mathbf{x}_t$ of user coordinates can be generated from a \emph{mixture} of a Brownian motion and a static process, denoted by $\mathbf{y}_t$ for which the conditional distribution $\mathbf{y}_{t+\Delta_t} \mid \mathbf{x}_{t}$ is given by
\[
\mathbf{y}_{t+\Delta t} \mid \mathbf{x}_{t} \sim
\begin{cases}
    \mathbf{x}_t + \sigma_{b_k} \sqrt{\Delta t} \cdot \boldsymbol{\epsilon} & \text{ with probability } 1-q_{b_k} \\
    \mathbf{x}_t & \text{ with probability } q_{b_k}
\end{cases}
\]
where $\sigma_{b_k}$ is a speed parameter specific to the building type, $\Delta t$ is a time step, $q_{b_k}$ is a probability of staying still specific to the building type, and $\boldsymbol{\epsilon} \sim \mathcal{N}(0, I)$ is a bivariate standard normal. We derive $\mathbf{x}_{t+\Delta_t} \mid \mathbf{x}_{t}$ from $\mathbf{y}_{t+\Delta t} \mid \mathbf{x}_{t}$ by conditioning on falling inside the building; that is
\[
\mathbf{x}_{t+\Delta t} \sim (\mathbf{y}_{t+\Delta t} \mid \mathbf{y}_{t+\Delta t} \in R_k),
\]
where $R_k$ is the region occupied by the current building $k$. Figure \ref{fig:constrained-brownian} depicts the conditional distribution of $\mathbf{x}_t$, which is used by the \texttt{sample\_step} method in the code. This method uses rejection sampling by sampling repeatedly from $\mathbf{y_t}$ until the point falls within $R_k$. A useful guideline in choosing the value of $\sigma_{b_k}$ is to set $\sigma_{b_k}$ equal to $z/1.96$, where $z$ is a realistic walking speed measured in blocks per time step---that is, there is a 95\% probability that the agent moves $z$ blocks in a time step of duration one minute. 

The heterogeneity in values for $q_{b_k}$ and $\sigma_{b_k}$ across buildings allows for different behaviors across building types. For example, a workplace may have a higher $q_{b_k}$ and lower $\sigma_{b_k}$ than a retail building, reflecting the idea that people tend to both move more and walk faster when shopping than when working. Moreover, our model provides additional flexibility in specifying different values for $q_{b_k}$ and $\sigma_{b_k}$ for the same building type across \textit{agents} to allow for heterogeneity in agent behavior. For example, there may be faster agents who stand still less frequently and move around faster compared to slower agents. To model movement \textit{between} buildings we modify the Brownian motion so that it is constrained to the street area, and we also add a drift ``towards'' $b_k$---implemented as the gradient of the shortest path under a one dimensional parametrization.

The agent’s trajectory is constructed by iteratively sampling steps with the \texttt{sample\_step} method, over the duration of each entry in the provided \texttt{destination\_diary}. Simultaneously, the agent’s ``ground-truth'' diary, in the variable \texttt{diary}, is updated by recording the time spent at each location or traveling between them (thus, the stop durations in \texttt{destination\_diary} are typically longer than the realized ones in \texttt{diary}).  

\subsection{Sparse timestamps sampled from a self-exciting point process}

To simulate realistic GPS trajectory data with sparse and clustered ping times (or exhibiting ``burstiness''), which are observed in commercial datasets, we make use of self-exciting point processes. These point processes have been used in diverse applications, including seismology, epidemiology, and crime modeling \cite{reinhart2018review}. The main idea in our implementation is to model two latent point processes, namely $N_s(t)$ and $N_e(t)$, that indicate when a burst of pings starts and ends, respectively. These latent events are used to define the conditional intensity of the process $N(t)$, which generates ping times either at a constant rate within a burst, or at a rate of zero outside of any burst (in a gap).

Formal definitions of the processes $N_s(t)$, $N_e(t)$, and $N(t)$ can be found in Appendix \ref{sec:app-nhpp}. In what follows we describe a simple procedure to sample from these processes given parameters $\beta_s$, $\beta_d$ and $\beta_p$, which respectively represent the average time in between burst starts, the average duration of a burst of pings, and the average time in between pings \emph{within} a burst. First, we generate ``burst start'' times by sampling inter-arrival times, $\{\Delta_i\}$, from an exponential with mean $\beta_{s}$. Second, for each burst start time $s_i$, we sample a ``burst end'' time $e_i$, by sampling a single inter-arrival time $d_i$---the burst duration---from an exponential with mean $\beta_d$, so that $e_i = s_i + d_i$. We are interested in modeling regimes in which bursts of pings are sparse in between long gaps, so, in the rare event in which the calculated end time of a burst extends beyond the start time of the next burst, we truncate. Third, we sample inter-arrival times, $\Delta_{t_{ij}}$, for the ping times $\{t_{ij}\}$ occurring between $s_i$ and $e_i$ from an exponential distribution with mean $\beta_p$. Thus, the ping times within burst $i$ are given by
\begin{align*}
    t_{ij} = s_i + \sum_{j=1}^{m} \Delta_{t_{ij}},
\end{align*}
\noindent where $t_{ij} \le e_i$. For practical purposes, we discretize the sampled ping times by rounding down to the nearest discrete time step.

Lastly, we introduce noise to the spatial coordinates of pings occurring at each of the times $\{t_{ij}\}$ which is meant to model the measurement error in GPS sensors. These errors--- caused by various factors like signal obstruction and atmospheric conditions---are often reported as metadata for each ping in the form of a ``horizontal accuracy'' parameter $\eta$, such that there is a 95\% probability that the true location is within $\eta$ meters of the measured location. 

We model this error by adding Gaussian noise to the true positions
\begin{align*}
\tilde{\mathbf{x}}_{t_{ij}} = \mathbf{x}_{t_{ij}} + \boldsymbol{\eta},
\end{align*}

\noindent where $\boldsymbol{\eta} \sim \mathcal{N}(0, \sigma_{\text{GPS}}^2 I)$, and $\sigma_{\text{GPS}}$ represents a standard deviation of the GPS error. To reconcile with the definition of ``horizontal accuracy'' used in commercial datasets, the value of $\sigma_{\text{GPS}}$ in our code is set to \texttt{ha}/1.96, where \texttt{ha} is the horizontal accuracy. Since our model's units are city blocks of side 15m, a value of $\sigma_{\text{GPS}}$ equal to $0.75 / 1.96$ will produce a ping with horizontal accuracy of 10 meters. 

\begin{figure}
\footnotesize
\begin{minipage}{0.08\textwidth}
\hspace{1pt}
\end{minipage}~
\begin{minipage}{0.8\textwidth}
    \input{snippet-sparsification}
\end{minipage}
    \caption{Example code that generates a sparse trajectory from a complete trajectory for an agent.}
    \label{fig:snippet-sparsification}
\end{figure}

\begin{figure}[!ht]
    \centering
    \includegraphics[width=\textwidth]{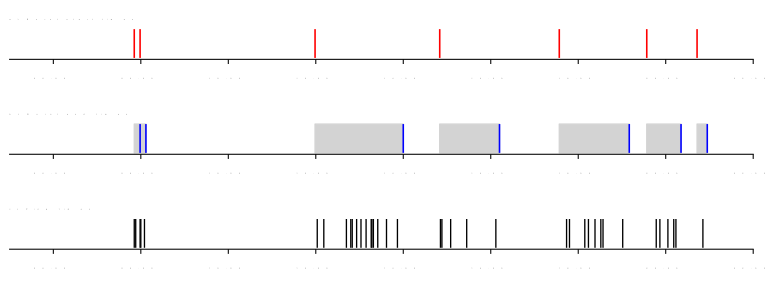}
    \caption{Example sparse signal over a 24-hour period for the sparse trajectory generated in Figure \ref{fig:snippet-sparsification}, together with latent variables. Panel \textbf{(a)} shows red lines at the burst start times generated by the latent process $N_s(t)$. Panel \textbf{(b)} shows blue lines at the burst end times generated by the latent process $N_e(t)$, as well as shaded areas representing each burst. Finally, Panel \textbf{(c)} shows timestamps of sampled pings. These ping times were simulated using the parameters \texttt{beta\_start=300}, \texttt{beta\_dur=60}, and \texttt{beta\_ping=10}.}
    \label{fig:nhpp}
\end{figure}

\section{Example 1: testing stop detection algorithms}
\label{sec:example1}

\begin{figure}[!ht]
    \centering
    \includegraphics[width=\linewidth]{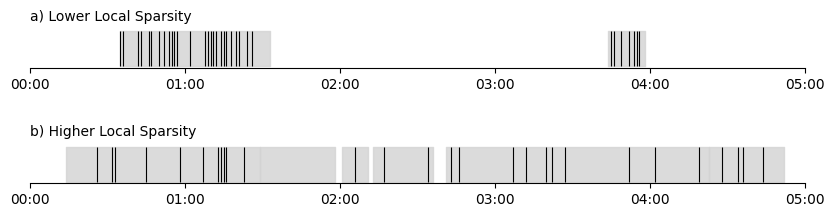}
    \caption{Vertical black lines indicate timestamps of sampled pings and shaded gray areas denote the duration of bursts generated for two levels of local sparsity: lower (\texttt{beta\_start=150}, \texttt{beta\_duration=20}, \texttt{beta\_ping=2}) or higher (\texttt{beta\_start=60}, \texttt{beta\_duration=40}, \texttt{beta\_ping=10})}
    \label{fig:sparsity-levels}
\end{figure}

As a demonstration of our model's capabilities, we adapt the robustness experiment proposed in \cite{barreras2024exciting}, which aimed to show that a common stop detection algorithm, DBScan, can be impacted by signal sparsity. In \cite{barreras2024exciting}, the notion of ``sparsity'' was captured only by the ping frequency of the overall trajectory, under a simple homogeneous process. In contrast, we can capture a more nuanced notion of sparsity by modeling ping times that are clustered into bursts. We demonstrate this with a toy example in which, starting from the same ``ground-truth'' trajectory, we can generate two sparse trajectories which have the same expected  same expected ping frequency for the whole trajectory (or ``global sparsity'') but have different ping frequencies within bursts of pings (or ``local sparsity''). The two levels of sparsity are visualized in Figure \ref{fig:sparsity-levels}.

\begin{figure}[!ht]
    \centering
    \includegraphics[width=0.8\linewidth]{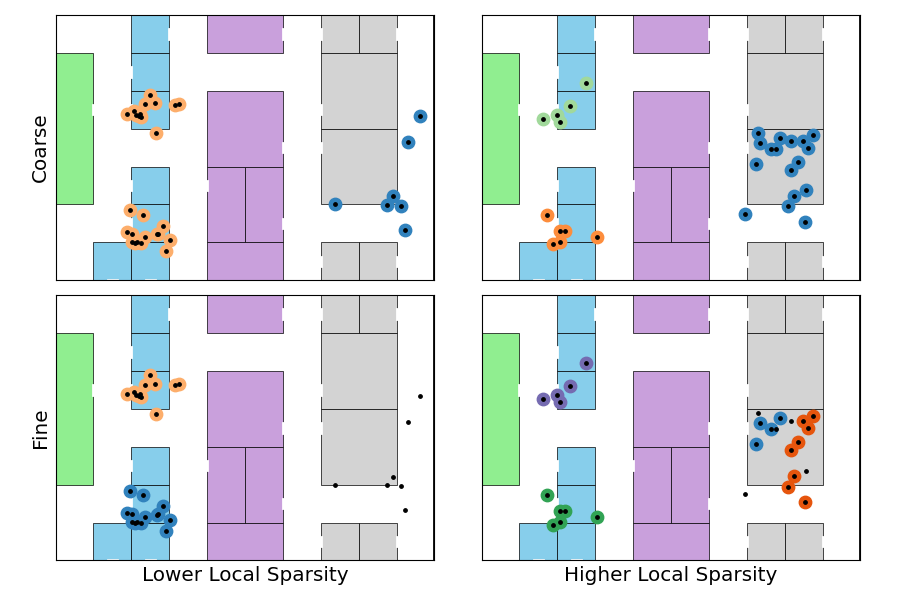}
    \caption{Stop detection results using  DBScan on simulated trajectories with varying local sparsity (higher vs. lower) and clustering resolution (fine vs. coarse). Colored points represent detected clusters, while uncolored points indicate noise.}
    \label{fig:dbscan}
\end{figure}

Figure \ref{fig:dbscan} depicts the results of applying two parameterizations of the stop detection algorithm, one ``fine'' and one ``coarse'', on those two trajectories. The specifics of the experiment are described in detail in the Appendix \ref{sec:app-example1}. The figure shows how changes in the clustering pattern of pings results in important variations of the output and can cause the same problems---such as omitting, merging, or splitting stops. This is a similar conclusion to that in \cite{barreras2024exciting}; however, in our example the overall expected ping frequency is kept at the same level. In the top left panel, we observe ``stop merging'', wherein a coarser parametrization struggles to differentiate stops at small neighboring establishments and instead clusters two stops as one. Conversely, a finer parametrization may completely miss a stop at a larger location (as in the bottom left panel) or erroneously cluster a larger stop as two distinct, shorter stops (as in the bottom right panel). Only the top right panel in this example accurately detected the three stops.

Our simple example suggests that ``global sparsity'' might be an inadequate metric to use for calibrating stop-detection algorithms---or for assessing the completeness of a dataset. Instead researchers could look into the clustered structure of pings in their datasets, the ``local'' pinging frequency, and the distribution of gaps in their data. 

\section{Example 2: The effect of noise on stops and dwell time}
\label{sec:example2}

We demonstrate our model's ability to fine-tune spatial variability in movement with an experiment examining how variance in movement and measurement noise affects stop detection algorithms. we simulate two users roaming a large location---the park in our sample city---with different movement settings. The first user moves slower and has a higher probability of staying still. Additionally, they have lower horizontal accuracy values (less noise), thus exhibiting lower spatial variance in their trajectory overall. The second user moves faster, is less likely to stay still, and has higher horizontal accuracy (more noise). 

\begin{figure}[!ht]
    \centering
    \includegraphics[width=0.8\linewidth]{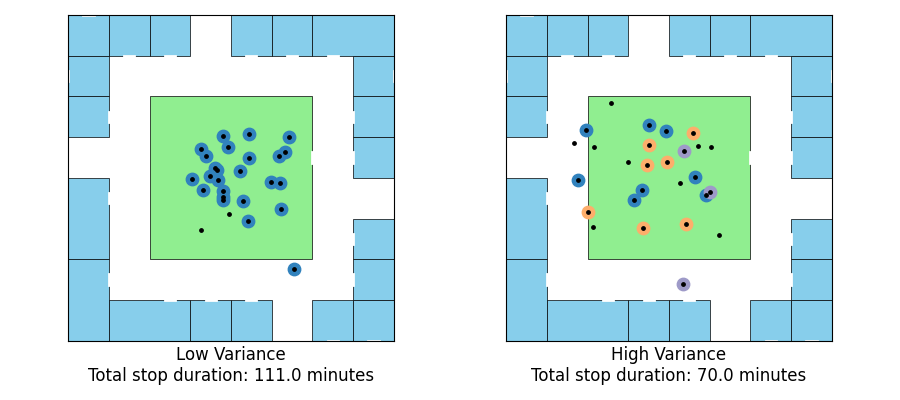}
    \caption{Stop detection results using the Project Lachesis algorithm \cite{hariharan2004project} on simulated trajectories with different levels of variance in agent movement and horizontal accuracy. Colored points represent detected clusters, while uncolored points indicate noise.}
    \label{fig:lachesis}
\end{figure}

We apply the sequential stop detection algorithm from \cite{hariharan2004project} to both trajectories. As shown in Figure \ref{fig:lachesis}, the user with low variance (left panel) accurately yields a single, continuous stop at the park, as expected. In contrast, the user with high variance (right panel) results in multiple fragmented stops due to increased noise and random movement, which disperse the pings and prevent the algorithm from clustering them correctly. This outcome suggests that algorithm parametrization should consider contextual features like the location, user speed, and horizontal accuracy for better results.

\section{Discussion and future work}

In this paper, we introduced Garden City, a mobility model and trajectory generator. Our model enables users to create---within a single sandbox environment---a stylized city, a population of agents with varied mobility profiles, and synthetic GPS trajectories that replicate features of real commercial GPS datasets. The primary purpose is to test processing algorithms against ``ground-truth'' data.

Our model is modular, offering great flexibility. Our code for generating cities allows users to create arbitrarily complex building layouts using building blocks and can be replaced with any compatible generator—for example, procedurally generated cities \cite{zufle2023urban} or adapted layouts from real cities \cite{tozluouglu2023synthetic}. The agent diary generator implements an EPR mobility model, but can be substituted with other mobility models or data-driven generators \cite{pappalardo2018data}. The GPS trajectory generator produces trajectories with tuneable features such as their temporal resolution, ``micro-level'' variability, spatial noise, and ping ``burstiness''---all of which can be tuned to accommodate different experiment needs.

The primary aim of our model is to facilitate testing the robustness and accuracy of human mobility data processing algorithms. Including a layer of ``micro-level'' randomness to the movement distinguishes our model from others in the literature, where agents may ``jump'' between tessellations \cite{pappalardo2018data}, remain stationary within buildings \cite{zufle2023urban}, or move rectilinearly between them \cite{tozluouglu2023synthetic}. Instead, our model allows fine-tuning of features that can introduce errors in algorithms. For example, increasing noise can reduce the likelihood that two pings are considered ``neighbors'' in a stop-detection algorithm; longer gaps without data can affect estimates of time spent at candidate home locations, impacting home-attribution algorithms; or varying user movement speeds can influence whether a stop in a larger location is detected or classified as movement between places.

Beyond this primary use case, the ability to generate datasets with the same format and features as commercial GPS datasets facilitates testing any code or pipeline involving such data before acquiring a commercial dataset license. For instance, users can employ Garden City to test and optimize code related to epidemic models using GPS data, such as those in \cite{chang2021mobility, pei2020differential}, among other applications.

Our current version of Garden City can be developed further in several interesting directions. One possibility is to generate city layouts using random models, allowing experiments that require larger-scale cities (such as those in epidemiology), avoiding limitations from an unrealistic default layout. As mentioned earlier, generating layouts based on real cities is also a promising avenue. Another direction is to implement other mobility models, both at the ``macro'' level—such as variations of the EPR model \cite{alessandretti2018evidence, pappalardo2016human}—and at the ``micro'' level—such as modeling different modes of transportation. There is potential in modeling trajectories in dense cities, where places of interest often share building plots or where the density of buildings increases signal noise. Additionally, calibrating the self-exciting point process we described to real data could guide the development of robustness experiments, focusing on sparsity regimes found in widely used datasets. The formulation as a self-exciting point process could facilitate this task, as there are analytical solutions to the likelihood of the data, allowing for Bayesian approaches like Markov Chain Monte Carlo (MCMC) to be used.

We hope that Garden City will be useful for many applications and can provide guidance on how to process real commercial GPS data.

\section{Code and data availability}

The Python code for Garden City is publicly available on GitHub at \url{https://github.com/Watts-Lab/nomad/}. The example Jupyter notebooks used to generate the outputs and plots in this report can be found in \url{https://github.com/Watts-Lab/nomad/tree/main/examples/garden-city}. Furthermore, example datasets generated with this methodology can be found in \url{https://github.com/Watts-Lab/nomad/tree/main/data}.

\section{Acknowledgments}
This work was possible with the support of the National Science Foundation, NSF award 2341932. 
\pagebreak

\bibliographystyle{plain}
\bibliography{references}

\begin{thebibliography}{10}

\bibitem{adjodah2021association}
Dhaval Adjodah, Karthik Dinakar, Matteo Chinazzi, Samuel~P Fraiberger, Alex Pentland, Samantha Bates, Kyle Staller, Alessandro Vespignani, and Deepak~L Bhatt.
\newblock Association between covid-19 outcomes and mask mandates, adherence, and attitudes.
\newblock {\em PloS one}, 16(6):e0252315, 2021.

\bibitem{alessandretti2022human}
Laura Alessandretti.
\newblock What human mobility data tell us about covid-19 spread.
\newblock {\em Nature Reviews Physics}, 4(1):12--13, 2022.

\bibitem{alessandretti2018evidence}
Laura Alessandretti, Piotr Sapiezynski, Vedran Sekara, Sune Lehmann, and Andrea Baronchelli.
\newblock Evidence for a conserved quantity in human mobility.
\newblock {\em Nature human behaviour}, 2(7):485--491, 2018.

\bibitem{aslak2020infostop}
Ulf Aslak and Laura Alessandretti.
\newblock Infostop: scalable stop-location detection in multi-user mobility data.
\newblock {\em arXiv preprint arXiv:2003.14370}, 2020.

\bibitem{barreras2024exciting}
Francisco Barreras and Duncan~J Watts.
\newblock The exciting potential and daunting challenge of using gps human-mobility data for epidemic modeling.
\newblock {\em Nature Computational Science}, pages 1--14, 2024.

\bibitem{becker2017literature}
Felix Becker and Kay~W Axhausen.
\newblock Literature review on surveys investigating the acceptance of automated vehicles.
\newblock {\em Transportation}, 44(6):1293--1306, 2017.

\bibitem{chang2021mobility}
Serina Chang, Emma Pierson, Pang~Wei Koh, Jaline Gerardin, Beth Redbird, David Grusky, and Jure Leskovec.
\newblock Mobility network models of covid-19 explain inequities and inform reopening.
\newblock {\em Nature}, 589(7840):82--87, 2021.

\bibitem{couture2021jue}
Victor Couture, Jonathan~I Dingel, Allison Green, Jessie Handbury, and Kevin~R Williams.
\newblock Jue insight: Measuring movement and social contact with smartphone data: a real-time application to covid-19.
\newblock {\em Journal of Urban Economics}, page 103328, 2021.

\bibitem{cresswell2006move}
Tim Cresswell.
\newblock {\em On the move: Mobility in the modern western world}.
\newblock Taylor \& Francis, 2006.

\bibitem{ester1996density}
Martin Ester, Hans-Peter Kriegel, J{\"o}rg Sander, Xiaowei Xu, et~al.
\newblock A density-based algorithm for discovering clusters in large spatial databases with noise.
\newblock In {\em kdd}, volume~96, pages 226--231, 1996.

\bibitem{gauvin2020gender}
Laetitia Gauvin, Michele Tizzoni, Simone Piaggesi, Andrew Young, Natalia Adler, Stefaan Verhulst, Leo Ferres, and Ciro Cattuto.
\newblock Gender gaps in urban mobility.
\newblock {\em Humanities and Social Sciences Communications}, 7(1):1--13, 2020.

\bibitem{hariharan2004project}
Ramaswamy Hariharan and Kentaro Toyama.
\newblock Project lachesis: parsing and modeling location histories.
\newblock In {\em International Conference on Geographic Information Science}, pages 106--124. Springer, 2004.

\bibitem{howard1898garden}
Ebenezer Howard.
\newblock {\em Garden cities of to-morrow}.
\newblock Swan Sonnenschein \& Co., 1898.

\bibitem{martin2023trackintel}
Henry Martin, Ye~Hong, Nina Wiedemann, Dominik Bucher, and Martin Raubal.
\newblock Trackintel: An open-source python library for human mobility analysis.
\newblock {\em Computers, Environment and Urban Systems}, 101:101938, 2023.

\bibitem{moro2021mobility}
Esteban Moro, Dan Calacci, Xiaowen Dong, and Alex Pentland.
\newblock Mobility patterns are associated with experienced income segregation in large us cities.
\newblock {\em Nature communications}, 12(1):4633, 2021.

\bibitem{painter2021political}
Marcus Painter and Tian Qiu.
\newblock Political beliefs affect compliance with government mandates.
\newblock {\em Journal of Economic Behavior \& Organization}, 185:688--701, 2021.

\bibitem{pappalardo2016human}
Luca Pappalardo, Salvatore Rinzivillo, and Filippo Simini.
\newblock Human mobility modelling: exploration and preferential return meet the gravity model.
\newblock {\em Procedia Computer Science}, 83:934--939, 2016.

\bibitem{pappalardo2018data}
Luca Pappalardo and Filippo Simini.
\newblock Data-driven generation of spatio-temporal routines in human mobility.
\newblock {\em Data Mining and Knowledge Discovery}, 32(3):787--829, 2018.

\bibitem{pappalardo2016analytical}
Luca Pappalardo, Maarten Vanhoof, Lorenzo Gabrielli, Zbigniew Smoreda, Dino Pedreschi, and Fosca Giannotti.
\newblock An analytical framework to nowcast well-being using mobile phone data.
\newblock {\em International Journal of Data Science and Analytics}, 2:75--92, 2016.

\bibitem{pei2020differential}
Sen Pei, Sasikiran Kandula, and Jeffrey Shaman.
\newblock Differential effects of intervention timing on covid-19 spread in the united states.
\newblock {\em Science advances}, 6(49):eabd6370, 2020.

\bibitem{pepe2020covid}
Emanuele Pepe, Paolo Bajardi, Laetitia Gauvin, Filippo Privitera, Brennan Lake, Ciro Cattuto, and Michele Tizzoni.
\newblock Covid-19 outbreak response, a dataset to assess mobility changes in italy following national lockdown.
\newblock {\em Scientific data}, 7(1):230, 2020.

\bibitem{reinhart2018review}
Alex Reinhart.
\newblock A review of self-exciting spatio-temporal point processes and their applications.
\newblock {\em Statistical Science}, 33(3):299--318, 2018.

\bibitem{ross2021household}
Stuart Ross, George Breckenridge, Mengdie Zhuang, and Ed~Manley.
\newblock Household visitation during the covid-19 pandemic.
\newblock {\em Scientific reports}, 11(1):22871, 2021.

\bibitem{song2010modelling}
Chaoming Song, Tal Koren, Pu~Wang, and Albert-L{\'a}szl{\'o} Barab{\'a}si.
\newblock Modelling the scaling properties of human mobility.
\newblock {\em Nature physics}, 6(10):818--823, 2010.

\bibitem{song2014prediction}
Xuan Song, Quanshi Zhang, Yoshihide Sekimoto, and Ryosuke Shibasaki.
\newblock Prediction of human emergency behavior and their mobility following large-scale disaster.
\newblock In {\em Proceedings of the 20th ACM SIGKDD international conference on Knowledge discovery and data mining}, pages 5--14, 2014.

\bibitem{tozluouglu2023synthetic}
{\c{C}}a{\u{g}}lar Tozluo{\u{g}}lu, Swapnil Dhamal, Sonia Yeh, Frances Sprei, Yuan Liao, Madhav Marathe, Christopher~L Barrett, and Devdatt Dubhashi.
\newblock A synthetic population of sweden: datasets of agents, households, and activity-travel patterns.
\newblock {\em Data in Brief}, 48:109209, 2023.

\bibitem{vanhoof2018assessing}
Maarten Vanhoof, Fernando Reis, Thomas Ploetz, and Zbigniew Smoreda.
\newblock Assessing the quality of home detection from mobile phone data for official statistics.
\newblock {\em Journal of official statistics}, 34(4):935--960, 2018.

\bibitem{yang2020tad}
Yuqing Yang, Jianghui Cai, Haifeng Yang, Jifu Zhang, and Xujun Zhao.
\newblock Tad: A trajectory clustering algorithm based on spatial-temporal density analysis.
\newblock {\em Expert Systems with Applications}, 139:112846, 2020.

\bibitem{zufle2023urban}
Andreas Z{\"u}fle, Carola Wenk, Dieter Pfoser, Andrew Crooks, Joon-Seok Kim, Hamdi Kavak, Umar Manzoor, and Hyunjee Jin.
\newblock Urban life: a model of people and places.
\newblock {\em Computational and Mathematical Organization Theory}, 29(1):20--51, 2023.

\end{thebibliography}

\clearpage
\appendix
\appendixsectionformat

\section{Self-exciting point process for generation of sparse pings} \label{sec:app-nhpp}

A simple point process, or counting process, is a stochastic process that can be defined as a counting measure $N( \cdot )$ on the real numbers, such that 
\begin{itemize}
    \item $N((-\infty, 0]) = 0$
    \item $P\left( N((a, b]) = n\right) = {\frac{(\Lambda_{a,b})^{n}}{n!}}e^{-\Lambda_{a,b}}$
    \item For disjoint intervals $I_1, \dots, I_k$, the random variables $N(I_1), \dots, N(I_k)$ are independent,
\end{itemize}

\noindent where $\Lambda_{a,b}$ is the average number of events expected in the interval $(a,b]$, assumed to be a right-continuous measure with a density $\lambda(t)$, called its \emph{intensity}, so that
\begin{align*}
    \Lambda_{a,b} = \int_{a}^b \lambda(t)\mathrm{d}t.
\end{align*}
\noindent Furthermore, we use the shorter notation $N(t)$ to denote the total number of events until time $t$, $N([0,t])$.

A point-process is called self-exciting (or a Hawkes process) if the intensity at time $t$ depends on the process history $\mathcal{H}_t$, with a linear dependence given by

\begin{align}
    \lambda(t \mid \mathcal{H}_t) &= \mu + \int_{0}^t g(t-u)\mathrm{d}N(u)\\ \label{eq:self-excite}
    &= \mu + \sum_{i:t_i<t} g(t-t_i),
\end{align}

\noindent where $\{t_i\}$ denote the event times in the history of the process, and $g$ is a \emph{triggering} function which typically models a decaying intensity over time. Self-exciting point processes can be naturally extended to multiple dimensions by making the conditional intensity in Equation \ref{eq:self-excite} dependent on the shared history of multiple processes.

The generative model used for generating sparse and clustered ping times is a self-exciting point process in three dimensions, consisting of the following three processes.

\begin{itemize}
    \item \textbf{Burst start times} are a homogeneous Poisson process $N_s\left( \cdot \right)$, with a constant intensity $\lambda_s$ given by
    \begin{align*}
        \lambda_s(t, \mathcal{H}_t) \coloneqq \dfrac{1}{\beta_{s}},
    \end{align*}
    \noindent with the parameter $\beta_{s}$ representing the average time between burst starts. We denote the event times with $s_i \in [0, T)$, such that $s_i < s_{i+1}$.
    
    \item \textbf{Burst end times} are a self-exciting point process $N_e\left( \cdot \right)$, with an intensity that is ``excited'' by a past event in the start time process $N_s$, and ``controlled'' by past events of $N_e$ itself---acting as an ``off switch''. The conditional intensity $\lambda_e$ is given by
    \begin{align}
        \lambda_e(t, \mathcal{H}_t) \coloneqq \dfrac{1}{\beta_{d}}\int_0^t \mathrm{d}N_s(u) - \dfrac{1}{\beta_{d}}\int_0^t \mathrm{d}N_e(v) \label{eq:end-times}
    \end{align}    
    If we denote these ``burst end'' event times with $e_i \in [0, T)$, such that $e_i < e_{i+1}$, then Equation \ref{eq:end-times} can be rewritten as
    \begin{align*}
        \lambda_e(t, \mathcal{H}_t) = \dfrac{1}{\beta_d}\left( N_s( t )-N_e( t )\right),
    \end{align*}    
    \noindent which makes it evident that the intensity of this process is $\dfrac{1}{\beta_d}$ when there is a new ``burst start'' event, and remains constant until there is a ``burst end '' event, which makes $N_s( t ) = N_e( t )$, and thus the intensity becomes zero (thus guaranteeing that the conditional intensity is never negative). It follows that $\beta_d$ is equal to the average burst duration.
    
    \item \textbf{Ping times} are a point process represented by $N(\cdot)$ which is also ``excited'' by a past event in the start time process $N_s$, and ``controlled'' by past events of $N_e$, but is independent of its own past, which generates ping times at a constant rate while the process is ``within'' a burst, and no pings otherwise. Formally, the conditional intensity is given by 
    \begin{align*}
        \lambda_p(t, \mathcal{H}_t) \coloneqq \dfrac{1}{\beta_{p}}\int_0^t \mathrm{d}N_s(u) - \dfrac{1}{\beta_{p}}\int_0^t \mathrm{d}N_e(v),
    \end{align*}
     \noindent which, again, can be written as 
    \begin{align*}
        \lambda_p(t, \mathcal{H}_t) = \dfrac{1}{\beta_p}\left( N_s( t )-N_e( t )\right),
    \end{align*}     
    
    \noindent thus, effectively representing times for a burst of pings in between burst starts and ends. 
\end{itemize}

\section{Model description and parameters for Example 1}
\label{sec:app-example1}

In line with \cite{barreras2024exciting}, we generate an agent with a simple trajectory consisting of consecutive 1-hour visits to two nearby homes followed by a 3-hour visit to a larger retail building. We then sample the agent's ``ground-truth'' trajectory at two levels of sparsity. The lower local sparsity sample expects a burst every 2.5 hours with a mean length of 20 minutes and a ping expected roughly every 2 minutes within the burst (\texttt{beta\_start=150}, \texttt{beta\_duration=20}, \texttt{beta\_ping=2}), for an expected total burst period of 40 minutes over 5 hours. The higher local sparsity sample expects a burst every hour with a mean length of 40 minutes and a ping expected roughly every 10 minutes within the burst (\texttt{beta\_start=60}, \texttt{beta\_duration=40}, \texttt{beta\_ping=10}), for an expected total burst period of 200 minutes over 5 hours. We deliberately choose these parameters to hold overall ping frequency fixed---both sparsity levels expect 20 pings over the 5-hour duration---in order to isolate the effect of the ``burstiness'' pattern. Figure \ref{fig:sparsity-levels} visually depicts the burst durations and sampled ping timestamps for the two levels of sparsity.

For the stop detection algorithm, as in \cite{barreras2024exciting}, we use a temporal variation of DBScan \cite{ester1996density}, that considers proximity in time to decide whether pings are ``neighbors''. At a broad level, the algorithm defines, for each point, a set of neighboring points which are within a distance threshold of \texttt{dist\_thresh} in space, and within time threshold of \texttt{time\_thresh} in time. Points with \texttt{min\_pts} neighbors or more are dubbed ``core points'' and clusters correspond to the connected components of the reachability graph between core points, incorporating non-core points to these clusters (but not reaching other points themselves). For each level of sparsity, we run two sets of DBSCAN parameters that affect the resolution of detected stops, dubbed ``coarse'' (\texttt{dist\_thresh = 2.25}, \texttt{time\_thresh = 120}, \texttt{min\_pts = 2}) and ``fine'' (\texttt{dist\_thresh = 1}, \texttt{time\_thresh = 45}, \texttt{min\_pts = 3}). The results of the experiment are shown in Figure \ref{fig:dbscan}.

\section{Model description and parameters for Example 2}
\label{sec:app-example2}

We simulate two agents with the same destination diary but different variance parameters. Both agents spend two hours at the park but one agent (Daniel) stays still with probability $q_\text{park}=0.75$ and moves around with a slower speed ($\sigma_{\text{park}} = 1.5/1.96$) whereas the other agent (Elaine) stays still with probability $q_\text{park}=0.1$ and moves around with a faster speed ($\sigma_{\text{park}} = 2.5/1.96$). As an additional source of variance, we sample Daniel's trajectory with a lower horizontal noise of $\eta=0.75$ and Elaine's trajectory with a higher horizontal noise of $\eta=1.5$. From the ``ground-truth'' trajectories, sparse trajectories are sampled with parameters \texttt{beta\_start=10}, \texttt{beta\_duration=1000}, \texttt{beta\_ping=5}, which is essentially one long burst that covers the entire two-hour stay during which a ping is expected every 5 minutes.

The clustering algorithm used is the Project Lachesis algorithm proposed by \cite{hariharan2004project}, which takes three parameters: \texttt{dur\_min}, \texttt{dt\_max}, and \texttt{delta\_roam}. The Lachesis algorithm greedily incorporates pings into a cluster as long as the cluster has a diameter of at most \texttt{delta\_roam}, no time gaps larger than \texttt{dt\_max}, and a duration larger than \texttt{dur\_min}. The parametrization applied to both agents' trajectories is \texttt{dur\_min=15}, \texttt{dt\_max=30}, and \texttt{delta\_roam=3}. The results of the experiment are shown in Figure \ref{fig:lachesis}.

\end{document}